\documentclass[a4paper,11pt]{article}
\usepackage[pdftex]{graphicx,xcolor}
\usepackage{float}
\usepackage{amsmath,amssymb,bm,amsthm}
\usepackage{ascmac}
\usepackage{array}
\usepackage{hyperref}
\usepackage[hang,small,bf]{caption}
\usepackage[subrefformat=parens]{subcaption}
\usepackage{natbib}
\usepackage{enumitem}

\bibpunct[:]{(}{)}{,}{a}{}{,}
\hypersetup{
	colorlinks=true,
	citecolor=blue,
	linkcolor=red,
	urlcolor=orange,
}
\title{Asymptotic Expansion for Nonlinear Filtering in the Small System Noise Regime}
\author{Masahiro Kurisaki\thanks{AIP Center, RIKEN\protect\\ email: \texttt{masahiro.kurisaki@riken.jp}}~\thanks{Japan Science and Technology Agency CREST \protect\\ The author was supported by Japan Science and Technology Agency CREST JP-MJCR2115, JSPS KAKENHI Grant Number JP24KJ0667, and RIKEN Special Postdoctoral Researcher Program.}}

\numberwithin{equation}{section}
\allowdisplaybreaks[3]

\theoremstyle{definition}

\theoremstyle{remark}

\begin{document}
\maketitle
\begin{abstract}
  We propose a new asymptotic expansion method for nonlinear filtering, based on a small parameter in the system noise. The conditional expectation is expanded as a power series in the noise level, with each coefficient computed by solving a system of ordinary differential equations. This approach mitigates the trade-off between computational efficiency and accuracy inherent in existing methods such as Gaussian approximations and particle filters. Moreover, by incorporating an Edgeworth-type expansion, our method captures complex features of the conditional distribution, such as multimodality, with significantly lower computational cost than conventional filtering algorithms.
\end{abstract}


\section{Introduction}\label{sec-intro}

Estimating hidden states from noisy observations is a fundamental problem in various fields, including signal processing, control theory, and finance. One classical framework for addressing this problem is filtering, which aims to estimate the hidden states of an unobserved stochastic process \(\{X_t\}_{t \geq 0}\) based on the observed data from a related stochastic process \(\{Y_t\}_{t \geq 0}\). Here, \(\{Y_t\}_{t \geq 0}\) provides partial information about \(\{X_t\}_{t \geq 0}\) through an observation model. 

The objective of the filtering problem is to compute the conditional expectation \(E[X_t|\mathcal{Y}_t]\) for every \(t \geq 0\), where \(\mathcal{Y}_t\) denotes the \(\sigma\)-field generated by \(\{Y_s\}_{0 \leq s \leq t}\). In the linear case, the filtering problem reduces to finite-dimensional equations, known as the Kalman filter \citep{kalman1960, kalman1961, Liptser2001,Liptser2001-2}.

On the other hand, in the general nonlinear case, the stochastic filtering problem is governed by the Kushner-Stratonovich equation \citep{stratonovich1960,kushner1964} or the equivalent Zakai equation \citep{Mortensen1966,Zakai1969,Duncan1970a,Duncan1970b}, which are generally known to be infinite-dimensional \citep{HAZEWINKEL1983331}. 

Since these equations are rarely tractable in practice due to the infinite-dimensionality, a wide variety of approximation methods have been developed, which may be broadly classified into three categories.

{\bf (i) Linearization or assumed-density methods} such as Extended Kalman filter \citep{picard1986,picard1991}, EnKF \citep{2003OcDyn..53..343E}, UKF \citep{882463}, Gaussian filter \citep{847749}, and projection filter \citep{Brigo1995,armstrong2019optimal} reduce the problem to a finite-dimensional system by locally linearizing the dynamics or projecting the solution onto a statistical manifold. While computationally efficient, these methods rely on strong structural assumptions and therefore offer limited theoretical guarantees regarding their accuracy.

{\bf (ii) PDE-based approaches}, such as finite difference \citep{Gyongy2003}, finite element \citep{Germani01021988} or spectral \citep{lototsky2011chaos} methods, attempt to solve the Zakai equation numerically. These offer rigorous approximations but suffer severely from the curse of dimensionality, making them impractical in high-dimensional systems.

{\bf (iii) Simulation-based methods}, including particle filters \citep{gordon1993, 4378823,6530707}, approximate the posterior distribution using sample-based methods and resampling. These are highly flexible and widely used, but require a large number of particles to ensure accuracy and often suffer from sample degeneracy.

These methods, despite their variety, ultimately face a fundamental trade-off: they either rely on strong approximations (such as linearization) to reduce computational complexity, or retain theoretical accuracy at the cost of prohibitively heavy computation.

In this paper, we propose a novel asymptotic expansion method for nonlinear filtering, designed to overcome the inherent trade-off between computational complexity and approximation accuracy present in existing approaches. Specifically, we introduce a small parameter $0<\epsilon<1$ and consider the model
\begin{align*}
  &dX_t^\epsilon = \alpha(t, X_t^\epsilon) \, dt + \epsilon \beta(t, X_t^\epsilon) \, dV_t,\\
  &dY_t^\epsilon = h(t, X_t^\epsilon) \, dt + \sigma(t) \, dW_t.
\end{align*}
The assumption of small system noise is both reasonable and commonly adopted in many practical applications, such as weather forecasting and ocean current analysis \citep{weather-forecast,DataAssimilationintheLowNoiseRegimewithApplicationtotheKuroshio}.

For this model, we expand the conditional expectation of the signal as a power series in $\epsilon$
\begin{align*}
  E[X_t^\epsilon|\mathcal{Y}_t^\epsilon]=m_t^{[0]}+m_t^{[1]}\epsilon+m_t^{[2]}\epsilon^2+\cdots,
\end{align*}
We demonstrate that the computation of each coefficient $m_t^{[i]}$ reduces to solving a system of (stochastic) ordinary differential equations that incorporate the observation process $Y$. This representation allows for significantly more efficient computation than PDE-based or simulation-based methods, while also offering improved accuracy and generality over linear approximation methods.

The structure of the paper is as follows. In Section~\ref{section-method}, we present our asymptotic expansion framework for nonlinear filtering. We begin by reducing the original nonlinear model to a polynomially perturbed linear model in Section~\ref{section-model-reduction}. In Section~\ref{section-expansion}, we derive the expansion of the conditional expectation for the reduced model, where each coefficient is expressed as an expectation under a newly defined probability measure. Section~\ref{section-recursive} shows that these coefficients satisfy (stochastic) ordinary differential equations driven by the observation process.

Section~\ref{section-numerical} provides numerical experiments that validate our method. In Section~\ref{section-numerical-linear}, we verify that the expansion converges to the exact filter in the linear case. Section~\ref{section-numerical-cubic} illustrates that our approach outperforms the Extended Kalman Filter for the cubic-sensor. In Section~\ref{section-numerical-density}, we further suggest that our method can be extended to asymptotically expand the conditional density in addition to the conditional expectation.

Finally, we summarize the advantages of our method in Section~\ref{section-discussion}.

  This work is based on Section 4 of our earlier paper [arXiv:2501.16333], but is reorganised to provide a self-contained account of the methodological aspects.

\section{Method}\label{section-method}
\subsection{Model Reduction}\label{section-model-reduction}
Let us consider a nonlinear unobserved process
\begin{align}
  \label{eq-theta}
  dX_t^\epsilon = \alpha(X_t^\epsilon) \, dt + \epsilon \, \beta(X_t^\epsilon) \, dV_t, 
  \quad X_0 = 0,
\end{align}
and an associated observed process
\begin{align}\label{eq-Y-eps}
  dY_t^\epsilon = h(X_t^\epsilon) \, dt + \sigma(t) \, dW_t, 
  \quad Y_0 = 0,
\end{align}
where $0 < \epsilon < 1$ is a constant, $\sigma(t)$ is a deterministic function, $\alpha$, $\beta$, and $h$ are nonlinear functions of class $C^\infty$, and $V$ and $W$ are independent Wiener processes. Also, we assume $X^\epsilon$, $Y^\epsilon$, $V$, and $W$ are one-dimensional for simplicity.

Our objective is to derive an asymptotic expansion of the conditional expectation $E[X_t^\epsilon  |  \mathcal{Y}_t^\epsilon]$ with respect to $\epsilon$, where $\{\mathcal{Y}_t^\epsilon\}_{t\geq 0}$ is the augmented filtration generated by $Y^\epsilon$ with null sets. To this end, we begin by expanding the process $X_t^\epsilon$ itself as a power series in $\epsilon$. For this purpose, let us write $\displaystyle \frac{\partial^i}{\partial \epsilon^i} X_t^\epsilon = X_t^{[i],\epsilon}$ and $X_t^{[i]} = X_t^{[i],\epsilon}$ for $i = 0, 1, 2, \cdots$. Then,  under regularity conditions, $X_t^{\epsilon}$ admits the Taylor expansion
\begin{align*}
  X_t^\epsilon=X_t^{[0]}+X_t^{[1]}\epsilon+\frac{1}{2}X_t^{[2]}\epsilon^2+\cdots
\end{align*}
and the derivatives $X_t^{[i]}~(i\geq 0)$ the following recursive differential equations:
\begin{align}
  dX_t^{[0]}=&\alpha(X_t^{[0]})dt,~~X_0^{[0]}=0,\\
  \label{eq-theta-1}dX_t^{[1]}=&\alpha'(X_t^{[0]})X_t^{[1]}dt+\beta(X_t^{[0]})dV_t,~~X_t^{[1]}=0,\\
  \label{eq-theta-2}dX_t^{[2]}=&\alpha'(X_t^{[0]})X_t^{[2]}dt+\alpha''(X_t^{[0]})(X_t^{[1]})^{2}dt+2\beta'(X_t^{[0]})X_t^{[1]}dV_t,~~X_t^{[2]}=0,\\
  &\vdots\nonumber
\end{align}
which is obtained by formally differentiating equation (\ref{eq-theta}) and evaluating it at $\epsilon=0$. For further details on such formal expansions, we refer the reader to \citet{bichteler1987malliavin}.

From these equations, we observe that $X_t^{[0]}$ is a deterministic process, and $X_t^{[1]}$ is a solution of the linear equation (\ref{eq-theta-1}). Furthermore, the equations for the higher-order derivatives take the form
\begin{align*}
  dX_t^{[k]}=\alpha'(X_t^{[0]})X_t^{[k]}dt+(\textrm{polynomials of } X_t^{[1]},\cdots,X_t^{[k-1]})\times (dt~\textrm{or}~dV_t).
\end{align*}
This implies that each $X^{[k]}~(k\geq 2)$ can be expressed recursively as a polynomial of lower-order derivatives, and ultimately as a polynomial (in the broad sense including integrals) in $X^{[1]}$. 

For example, if we assume $\beta(X_t^{[0]})\neq 0$ for every $t\geq 0$, then (\ref{eq-theta-1}) and (\ref{eq-theta-2}) yields
\begin{align*}
  X_t^{[2]}=&\int_0^t \exp\left( \int_s^t \alpha'(X_u^{[0]})du \right)\alpha''(X_s^{[0]})(X_s^{[1]})^{2}ds\\
  &+2\int_0^t \exp\left( \int_s^t \alpha'(X_u^{[0]})du \right)\beta'(X_t^{[0]})X_t^{[1]}dV_s\\
  =&\int_0^t \exp\left( \int_s^t \alpha'(X_u^{[0]})du \right)\alpha''(X_s^{[0]})(X_s^{[1]})^{2}ds\\
  &+2\int_0^t \exp\left( \int_s^t \alpha'(X_u^{[0]})du \right)\beta'(X_t^{[0]})X_t^{[1]}\frac{dX_t^{[1]}-\alpha'(X_t^{[0]})dt}{\beta(X_t^{[0]})}.
\end{align*}
In virtue of It\^o's formula, this can be transformed into
\begin{align}
  \begin{split}
    &X_t^{[2]}=\frac{b'(X_t^{[0]})}{b(X_t^{[0]})}(X_t^{[1]})^2\\
  &+\int_0^t \exp\left( \int_s^t \alpha'(X_u^{[0]})du \right) \left\{ a''(X_s^{[0]})-\frac{\alpha'(X_s^{[0]})\beta'(X_s^{[0]})+\beta''(X_s^{[0]})}{\beta(X_s^{[0]})}+\frac{\beta'(X_s^{[0]})}{\beta(X_s^{[0]})} \right\}(X_s^{[1]})^2ds\\
  &-\int_0^t \exp\left( \int_s^t \alpha'(X_u^{[0]})du \right)\beta'(X_s^{[0]})\beta(X_s^{[0]})ds,
  \end{split}\label{eq-X_t2}
\end{align}
which is written by $(X^{[1]})^2$ and its integral.

Given this polynomial structure of the higher-order derivatives, the variable $h(X_t^\epsilon)$ in equation \eqref{eq-Y-eps} admits the expansion
\begin{align}
  h(X_t^{\epsilon})=&h(X_t^{[0]})+h'(X_t^{[0]})X_t^{[1]}\epsilon\nonumber\\
  &+\frac{1}{2}\left\{ h''(X_t^{[0]})(X_t^{[1]})^2+h'(X_t^{[0]})X_t^{[2]} \right\}\epsilon^2+\cdots\nonumber\\
  =&h(X_t^{[0]})+h'(X_t^{[0]})X_t^{[1]}\epsilon+\sum_{i=2}^\infty (\textrm{polynomial of }X^{[1]})\epsilon^i.\label{eq-expansion-h}
\end{align}
Thus, the original system given by equations~\eqref{eq-theta} and~\eqref{eq-Y-eps} can be reformulated as a polynomially perturbed linear model:
\begin{align*}
  &dX_t^{[1]}=\alpha'(X_t^{[0]})X_t^{[1]}dt+\beta(X_t^{[0]})dV_t,\\
  &dY_t^\epsilon=\left\{ h(X_t^{[0]})+h'(X_t^{[0]})X_t^{[1]}\epsilon+\sum_{i=2}^\infty (\textrm{polynomial of }X^{[1]})\epsilon^i \right\}dt+\sigma(t)dW_t.
\end{align*}
Correspondingly, the conditional expectation admits the expansion
\begin{align*}
  E[X_t^\epsilon|\mathcal{Y}_t^\epsilon]&=X_t^{[0]}+E[X_t^{[1]}|\mathcal{Y}_t^\epsilon]\epsilon+\frac{1}{2}E[X_t^{[2]}|\mathcal{Y}_t^\epsilon]\epsilon^2+\cdots\\
  &=X_t^{[0]}+E[X_t^{[1]}|\mathcal{Y}_t^\epsilon]\epsilon+\sum_{i=2}^\infty E\bigl[(\textrm{polynomial of }X^{[1]})|\mathcal{Y}_t^\epsilon\bigr]\epsilon^i.
\end{align*}

Thus, let us rewrite $X^{[1]}$ as $X$, and consider the simplified model
\begin{alignat}{2}
  \label{eq-model-sim1}
    & dX_t = aX_t \, dt + b \, dV_t, \qquad & X_0 = 0, \\
  \label{eq-model-sim2}
    & dY_t^\epsilon = (cX_t + \epsilon g(X_t)) \, dt + \sigma \, dW_t, \qquad & Y_0^\epsilon = 0,
\end{alignat}
where $a, b, c \in \mathbb{R}$, $0 < \epsilon < 1$, $\sigma > 0$, and $g$ is a polynomial. 

The perturbation term $\epsilon g(X_t)$ in the original model corresponds to the power series $\displaystyle \sum_{i=2}^\infty (\textrm{polynomial of }X^{[1]})\epsilon^i$ that arises in the original model. Each such "polynomial" may include integral expressions with multiplicative kernels such as $\exp\left( \int_s^t \alpha'(X_u^{[0]})du \right)$ are included, as seen in equation~\eqref{eq-X_t2}. However, for the purpose of illustrating the core idea of our proposed algorithm, it is sufficient to consider the reduced model given by equations~\eqref{eq-model-sim1} and~\eqref{eq-model-sim2}.\\
{\bf Remark:} In deriving equation~\eqref{eq-X_t2}, we assumed that (\ref{eq-X_t2}), we assumed that $\beta(X_t^{[0]}) \neq 0$ for all $t$. In more general settings---such as when $\beta(X_t^{[0]})$ vanishes at some points, or when, $X$ and $Y$ are multi-dimensional---the model reduction step can still be carried out by augmenting the state process to include both $X^{[1]}$ and the driving noise $V$, i.e., by defining the extended process $\tilde{X}_t=(X_t^{[1]},V_t)$. In such cases, the “polynomial” terms may involve stochastic integrals with respect to $d\tilde{X}_t$, but the subsequent steps of the algorithm remain valid with appropriate modifications.

\subsection{Expansion of the conditional expectation}\label{section-expansion}
Now, we consider the conditional expectation $E[f(X_t) | \mathcal{Y}_t^\epsilon]$ under the model given by equations~\eqref{eq-model-sim1} and~\eqref{eq-model-sim2}, where $f$ is a polynomial. To simplify the presentation of the algorithm, we focus on the case $f(x)=x$.

Fix $T > 0$, and introduce a new probability measure $Q$ by
\begin{align*}
  &Q(A) = E\left[ 1_A \exp\left( 
    -\int_0^T \frac{cX_t + \epsilon g(X_t)}{\sigma^2} \, dW_t 
    - \frac{1}{2} \int_0^T \frac{\{cX_t + \epsilon g(X_t)\}^2}{\sigma^2} \, dt 
  \right) \right]
\end{align*}
for $A \in \mathcal{F}_T$. Then, by Bayes' formula—--also known as the Kallianpur-Striebel formula \citep{bain2009fundamentals}---the conditional expectation can be expressed as
\begin{align}
  \label{eq4-1}
  \begin{split}
    &E[X_t |  \mathcal{Y}_t^\epsilon] = \frac{\displaystyle E_Q\left[ 
      X_t \exp\left( 
        \int_0^t \frac{cX_s + \epsilon g(X_s)}{\sigma^2} \, dY_s^\epsilon 
        - \frac{1}{2} \int_0^t \frac{\{cX_s + \epsilon g(X_s)\}^2}{\sigma^2} \, ds 
      \right) 
       \middle|  \mathcal{Y}_t^\epsilon 
    \right]}{\displaystyle E_Q\left[ 
      \exp\left( 
        \int_0^t \frac{cX_s + \epsilon g(X_s)}{\sigma^2} \, dY_s^\epsilon 
        - \frac{1}{2} \int_0^t \frac{\{cX_s + \epsilon g(X_s)\}^2}{\sigma^2} \, ds 
      \right) 
       \middle|  \mathcal{Y}_t^\epsilon
    \right]}:=m_t^\epsilon
  \end{split}
\end{align}
for $0 \leq t \leq T$, where $E_Q$ denotes the expectation with respect to $Q$. 

Let us consider expanding $m_t^\epsilon$ with respect to $\epsilon$. The coefficient of $\epsilon^0$ in the expansion of $m_t^\epsilon$ can be written as
\begin{align*}
  m_t^0=\frac{\displaystyle E_Q\left[ X_t \exp\left( \int_0^t \frac{cX_s}{\sigma^2}dY_s^\epsilon-\frac{1}{2}\int_0^t \frac{(cX_s)^2}{\sigma^2}ds \right)\middle|\mathcal{Y}_t^\epsilon \right]}{\displaystyle E_Q\left[  \exp\left( \int_0^t \frac{cX_s}{\sigma^2}dY_s^\epsilon-\frac{1}{2}\int_0^t \frac{(cX_s)^2}{\sigma^2}ds \right)\middle|\mathcal{Y}_t^\epsilon \right]}=\tilde{E}_t[X_t],
\end{align*}
where
\begin{align}
  \label{eq-def-tilde-E}\tilde{E}_t[U]=\frac{\displaystyle E_Q\left[ U\exp\left( \int_0^t \frac{cX_s}{\sigma^2}dY_s^\epsilon-\frac{1}{2}\int_0^t \frac{(cX_s)^2}{\sigma^2}ds \right)\middle|\mathcal{Y}_t^\epsilon \right]}{\displaystyle E_Q\left[ \exp\left( \int_0^t \frac{cX_s}{\sigma^2}dY_s^\epsilon-\frac{1}{2}\int_0^t \frac{(cX_s)^2}{\sigma^2}ds \right)\middle|\mathcal{Y}_t^\epsilon \right]}.
\end{align}
Furthermore, differentiating (\ref{eq4-1}) yields
\begin{align*}
  &\left.\frac{d}{d\epsilon}m_t^\epsilon\right|_{\epsilon=0}\\
  =&\frac{\displaystyle E_Q\left[ X_t \int_0^t g(X_s)(dY_s^\epsilon-cX_sds) \exp\left( \int_0^t \frac{cX_s}{\sigma^2}dY_s^\epsilon-\frac{1}{2}\int_0^t \frac{(cX_s)^2}{\sigma^2}ds \right)\middle|\mathcal{Y}_t^\epsilon \right]}{\displaystyle E_Q\left[  \exp\left( \int_0^t \frac{cX_s}{\sigma^2}dY_s^\epsilon-\frac{1}{2}\int_0^t \frac{(cX_s)^2}{\sigma^2}ds \right)\middle|\mathcal{Y}_t^\epsilon \right]}\\
  &-\frac{\displaystyle E_Q\left[ X_t  \exp\left( \int_0^t \frac{cX_s}{\sigma^2}dY_s^\epsilon-\frac{1}{2}\int_0^t \frac{(cX_s)^2}{\sigma^2}ds \right)\middle|\mathcal{Y}_t^\epsilon \right]}{\displaystyle E_Q\left[  \exp\left( \int_0^t \frac{cX_s}{\sigma^2}dY_s^\epsilon-\frac{1}{2}\int_0^t \frac{(cX_s)^2}{\sigma^2}ds \right)\middle|\mathcal{Y}_t^\epsilon \right]}\\
  &\times \frac{\displaystyle E_Q\left[  \int_0^t g(X_s)(dY_s^\epsilon-cX_sds) \exp\left( \int_0^t \frac{cX_s}{\sigma^2}dY_s^\epsilon-\frac{1}{2}\int_0^t \frac{(cX_s)^2}{\sigma^2}ds \right)\middle|\mathcal{Y}_t^\epsilon \right]}{\displaystyle E_Q\left[  \exp\left( \int_0^t \frac{cX_s}{\sigma^2}dY_s^\epsilon-\frac{1}{2}\int_0^t \frac{(cX_s)^2}{\sigma^2}ds \right)\middle|\mathcal{Y}_t^\epsilon \right]}\\
  =&\tilde{E}_t\left[ X_t\int_0^t g(X_s)(dY_s^\epsilon-cX_sds) \right]-\tilde{E}_t\left[ X_t \right]\tilde{E}_t\left[ \int_0^t g(X_s)(dY_s^\epsilon-cX_sds) \right],
\end{align*}
which gives the coefficient of $\epsilon^1$. Here, note that \( Y^\epsilon \) is the given observation and is treated as fixed; in particular, it is not differentiated.

In order to find higher order coefficients, write
\begin{align*}
  &\exp\left( \int_0^t \frac{cX_s+\epsilon g(X_s)}{\sigma^2}dY_s^\epsilon-\frac{1}{2}\int_0^t\frac{\{cX_s+\epsilon g(X_s)\}^2}{\sigma^2} ds \right)\\
  =& \exp\left( \int_0^t \frac{cX_s}{\sigma^2}dY_s^\epsilon-\frac{1}{2}\int_0^t \frac{(cX_s)^2}{\sigma^2}ds \right)\exp\left( \epsilon \int_0^t \frac{g(X_s)}{\sigma^2}(dY_s^\epsilon-cX_sds)-\frac{1}{2}\epsilon^2\int_0^t \frac{g(X_s)^2}{\sigma^2}ds \right).
\end{align*}
Here, set
\begin{align*}
  K_t^\epsilon=\exp\left( \epsilon \int_0^t \frac{g(X_s)}{\sigma^2}(dY_s^\epsilon-cX_sds)-\frac{1}{2}\epsilon^2\int_0^t \frac{g(X_s)^2}{\sigma^2}ds \right).
\end{align*}
Then $K_t^\epsilon$ is an exponential local martingale, and we have the expansion
\begin{align*}
  K_t^\epsilon=&1+\frac{\epsilon}{\sigma^2}\int_0^t K_s^\epsilon g(X_s)(dY_s^\epsilon-cX_sds)\\
  =&1+\frac{\epsilon}{\sigma^2}\int_0^t g(X_s)(dY_s^\epsilon-cX_sds)\\
  &+\frac{\epsilon^2}{\sigma^4}\int_0^t K_u^\epsilon g(X_u)(dY_u^\epsilon-cX_udu)g(X_s)(dY_s^\epsilon-cX_sds)\\
  =&\cdots\\
  =&1+\sum_{i=1}^n\frac{\epsilon^i}{\sigma^{2i}}\int_0^t\int_0^{t_i}\cdots \int_0^{t_2}g(X_{t_1})\cdots g(X_{t_{n}})\\
  &\times (dY_{t_1}^\epsilon-cX_{t_1}dt_1)\cdots (dY_{t_i}^\epsilon-cX_{t_i}dt_i)\\
  &+\frac{\epsilon^{n+1}}{\sigma^{2(n+1)}}\int_0^t\int_0^{t_{n+1}}\cdots \int_0^{t_2}K_{t_1}^\epsilon g(X_{t_1})\cdots g(X_{t_{n+1}})\\
  &\times (dY_{t_1}^\epsilon-cX_{t_1}dt_1)\cdots (dY_{t_{n+1}}^\epsilon-cX_{t_{n+1}}dt_{n+1})
\end{align*}
for any $n \in \mathbb{N}$. The last term is shown to be $O(\epsilon^{n+1})$ in some sense.
Therefore, (\ref{eq4-1}) can be expanded as
\begin{align}
  E[X_t|\mathcal{Y}_t^\epsilon]=&\frac{\displaystyle E_Q\left[ X_t K_t^\epsilon\exp\left( \int_0^t \frac{cX_s}{\sigma^2}dY_s^\epsilon-\frac{1}{2}\int_0^t \frac{(cX_s)^2}{\sigma^2}ds \right)\middle|\mathcal{Y}_t^\epsilon \right]}{\displaystyle E_Q\left[ K_t^\epsilon\exp\left( \int_0^t \frac{cX_s}{\sigma^2}dY_s^\epsilon-\frac{1}{2}\int_0^t \frac{(cX_s)^2}{\sigma^2}ds \right)\middle|\mathcal{Y}_t^\epsilon \right]}\nonumber\\
  =&\frac{\displaystyle \frac{\displaystyle E_Q\left[ X_t K_t^\epsilon\exp\left( \int_0^t \frac{cX_s}{\sigma^2}dY_s^\epsilon-\frac{1}{2}\int_0^t \frac{(cX_s)^2}{\sigma^2}ds \right)\middle|\mathcal{Y}_t^\epsilon \right]}{\displaystyle E_Q\left[ \exp\left( \int_0^t \frac{cX_s}{\sigma^2}dY_s^\epsilon-\frac{1}{2}\int_0^t \frac{(cX_s)^2}{\sigma^2}ds \right)\middle|\mathcal{Y}_t^\epsilon \right]}}{\displaystyle \frac{\displaystyle E_Q\left[ K_t^\epsilon\exp\left( \int_0^t \frac{cX_s}{\sigma^2}dY_s^\epsilon-\frac{1}{2}\int_0^t \frac{(cX_s)^2}{\sigma^2}ds \right)\middle|\mathcal{Y}_t^\epsilon \right]}{\displaystyle E_Q\left[ \exp\left( \int_0^t \frac{cX_s}{\sigma^2}dY_s^\epsilon-\frac{1}{2}\int_0^t \frac{(cX_s)^2}{\sigma^2}ds \right)\middle|\mathcal{Y}_t^\epsilon \right]}}\nonumber\\
  \label{eq4-5}\begin{split}
    =&\frac{\tilde{E}_t[X_tK_t^\epsilon]}{\tilde{E}_t[K_t^\epsilon]}\\
  =&J_t^0(X_t) +\frac{1}{\sigma^{2}} \left\{ J_t^1(X_t) - J_t^0(X_t) J_t^1(1) \right\} \epsilon \\
  &+ \frac{1}{\sigma^{4}}\left\{ J_t^2(X_t) - J_t^0(X_t) J_t^2(1) - J_t^1(X_t) J_t^1(1) + J_t^0(X_t) J_t^1(1)^2 \right\} \epsilon^2 \\
  &+ \cdots + \frac{1}{\sigma^{2n}}\sum_{k=0}^n \sum_{p=1}^{n-k} \sum_{\substack{i_1,\cdots,i_p \in \mathbb{N} \\ i_1+\cdots+i_p=n-k}} (-1)^p J_t^k(X_t) J_t^{i_1}(1) \cdots J_t^{i_p}(1) \epsilon^n\\
  &+O(\epsilon^{n+1}),
  \end{split}  
\end{align}
where
\begin{align}
  \label{eq4-3}\begin{split}
    J_t^i(U) =  \tilde{E}_t\Biggl[ &U \int_0^t \int_0^{t_i} \cdots \int_0^{t_2} g(X_{t_1}) \cdots g(X_{t_i})  (dY_{t_1}^\epsilon - cX_{t_1}  dt_1) \cdots (dY_{t_i}^\epsilon - cX_{t_i}  dt_i) \Biggr]
  \end{split}    
\end{align}
for a random variable $U$. Thus, the expansion of $E[X_t|\mathcal{Y}_t^\epsilon]$ is reduced to computing the (conditional) expectation (\ref{eq4-3}).

\subsection{Recursive formulae for the expectation under $\tilde{E}_t$}\label{section-recursive}
To compute the expectation in equation~\eqref{eq4-3}, we analyze the distribution of the process $\{X_s\}_{0\leq s\leq t}$ under the modified expectation $\tilde{E}_t$. The definition of $\tilde{E}_t$ is given in equation~\eqref{eq-def-tilde-E}, and it corresponds to the normalized conditional expectation obtained by setting $g(X_t)=0$ in the original formula~\eqref{eq4-1}. In other words, the law of $\{X_s\}_{0\leq s\leq t}$ under $\tilde{E}_t$ is the conditional distribution of $\{X_s\}_{0\leq s\leq t}$ given $\mathcal{Y}_t^\epsilon$ in the linear system
\begin{alignat*}{2}
    & dX_t = aX_t \, dt + b \, dV_t, \qquad & X_0 = 0, \\
    & dY_t^\epsilon = cX_t  \, dt + \sigma \, dW_t, \qquad & Y_0^\epsilon = 0.
\end{alignat*}

For this linear Gaussian system, it is well known that $\{X_s\}_{0\leq s\leq t}$ is conditionally Gaussian given the observation $\mathcal{Y}_t^\epsilon$. Since $g$ is a polynomial function, the expectation in~\eqref{eq4-3} can be expressed in terms of the conditional mean and covariance of $X$ under $\tilde{E}_t$, denoted by
\begin{align*}
 \mu_{s;t}=\tilde{E}_t[X_s],~~\mathrm{Cov}_{\tilde{E}_t}(X_{s},X_{u})=\gamma(s,u;t). 
\end{align*}
For instance, when $g(x)=x^3$, we have
\begin{align*}
  J_t^0(X_t)=&\tilde{E}_t[X_t]=\mu_{t;t},\\
  J_t^1(X_t)=&\tilde{E}_t\left[ X_t\int_0^t X_s^3 (dY_s^\epsilon - cX_sds) \right]\nonumber\\
  =&\int_0^t \left( \tilde{E}_t\left[ X_tX_s^3 \right] dY_s^\epsilon - c\tilde{E}_t\left[ X_tX_s^4 \right]ds\right)\nonumber\\
    =&\mu_{t;t}\int_0^t \mu_{s;t}^3 (dY_s^\epsilon-c\mu_{s;t}ds)
    +3\mu_{t;t}\int_0^t \mu_{s;t}\gamma(s,s;t) (dY_s^\epsilon-c\mu_{s;t}ds)\\
  &+3\int_0^t \mu_{s;t}^2 \gamma(s,t;t) (dY_s^\epsilon-c\mu_{s;t}ds)\\
  &-c\int_0^t \mu_{s;t}^3 \gamma(s,t;t)ds
  -3c\mu_{t;t}\int_0^t \mu_{s;t}^2 \gamma(s,t;t)ds\\
  &+3\int_0^t \gamma(s,t;t)\gamma(s,s;t) (dY_s^\epsilon-c\mu_{s;t}ds)\\
  &-9c\int_0^t \mu_{s;t}\gamma(s,t;t)\gamma(s,s;t)ds-3c\mu_{t;t}\int_0^t \gamma(s,t;t)\gamma(s,s;t)ds,\\
  J_t^0(1)=&\tilde{E}_t\left[ \int_0^t X_s^3 (dY_s^\epsilon - cX_s \, ds) \right]\nonumber\\
  =&\int_0^t \left( \tilde{E}_t\left[ X_s^3 \right] dY_s^\epsilon - c\tilde{E}_t\left[ X_s^4 \right]ds\right)\nonumber\\
    =&\int_0^t \mu_{s;t}^3 (dY_s^\epsilon-c\mu_{s;t}ds)+3\int_0^t \mu_{s;t}\gamma(s,s;t) (dY_s^\epsilon-c\mu_{s;t}ds)\\
  &-3c\int_0^t \mu_{s;t}^2 \gamma(s,t;t)ds-3c\int_0^t \gamma(s,t;t)\gamma(s,s;t)ds.
\end{align*}
Using these expressions, the coefficient of $\epsilon^1$ in (\ref{eq4-5}) is represented as 
\begin{align}
  &J_t^1(X_t) - J_t^0(X_t) J_t^1(1) \nonumber \\
  \label{eq4-6}\begin{split}
    =& 3\int_0^t \mu_{s;t}^2 \gamma(s,t;t) (dY_s^\epsilon - c\mu_{s;t} ds)
      - c \int_0^t \mu_{s;t}^3 \gamma(s,t;t) \, ds \\
      &+ 3\int_0^t \gamma(s,t;t) \gamma(s,s;t) (dY_s^\epsilon - c\mu_{s;t}ds) - 9c \int_0^t \mu_{s;t} \gamma(s,t;t) \gamma(s,s;t) \, ds.
  \end{split}      
\end{align}
Here, for numerical stability, we preserved $dY_s^\epsilon - c\mu_{s;t} \, ds$ without fully expanding integrals. Note also that some of the stochastic integrals above are not well-defined, since $\{\mu_{s;t}\}_{0 \leq s \leq t}$ are not adapted (they depend on the full history of $\{Y_s^\epsilon\}_{0 \leq s \leq t}$). However, we can justify the procedure by appropriately interpreting the integrals.

In any case, $J_t^i(X_t)$ and $J_t^i(1)$ are expanded into a sum of terms of the form
\begin{align}
  \label{eq-Avec}\begin{split}
    A_t(p,q,r,\alpha;n)
    =&\int_0^t\int_0^{t_n}\cdots \int_0^{t_{2}}\prod_{i=1}^n(\mu_{t_i;t})^{p_i}\prod_{i=1}^n \mathrm{Cov}(\xi_{t;t},\xi_{t_i;t})^{q_{i}}\\
    &\times \prod_{1\leq i\leq j\leq n}\mathrm{Cov}(\xi_{t_i;t},\xi_{t_j;t})^{r_{ij}}d\lambda_{t_1;t}^{(\alpha_1)}\cdots d\lambda_{t_n;t}^{(\alpha_n)},
  \end{split}  
\end{align}
where $n \in \mathbb{N}$, $p=(p_i)_{i=1,\cdots,n} \in \mathbb{Z}_+^n,~q=(q_i)_{i=1,\cdots,n} \in \mathbb{Z}_+^n$, $r=(r_{ij})_{i,j=1,\cdots,n} \in M_n(\mathbb{Z}_+)$, $\alpha=(\alpha_i)_{i=1,\cdots,n} \in \{0,1\}^n$, and
\begin{align*}
  d\lambda_{s;t}^{(i)}=\begin{cases}
    ds&(i=0)\\
    dY_s^\epsilon-c\mu_{s;t}ds&(i=1).
  \end{cases}
\end{align*}
For example, (\ref{eq4-6}) can be written as
\begin{align*}
  3A_t(2,1,0,1;1)-cA_t(3,1,0,0;1)+3A_t(0,1,1,1;1)-6cA_t(1,1,1,0;1).
\end{align*}
To derive a recursive formula for $A_t(p,q,r,\alpha;n)$, we use the following recursive equations:
\begin{align}
  \label{eq4-7}&d_t\mu_{t;t}=a\mu_{t;t}dt+ \frac{c\gamma(t)}{\sigma^2}(dY_t^\epsilon-c\mu_{t;t}dt),\\  
  &\label{eq4-8}d_t\mu_{s;t}=\frac{c}{\sigma^2}\gamma(s,t;t)(dY_t^\epsilon-c\mu_{t;t}dt),\\
  \label{eq4-9}&d_t\gamma(s,t;t)=\left\{ a-\frac{c^2\gamma(t)}{\sigma^2} \right\}\gamma(s,t;t)dt,\\
  \label{eq4-10}&d_t\gamma(s,u;t)=-\frac{c^2}{\sigma^2}\gamma(s,t;t)
    \gamma(u,t;t)dt,
\end{align}
where $\gamma(t)=\gamma(t,t;t)$ is the solution of
\begin{align}
  \label{eq4-11}\frac{d}{dt}\gamma(t)=-\frac{c^2}{\sigma^2}\gamma(t)^2+2a\gamma(t)+b^2.
\end{align}
Here, (\ref{eq4-7}) and (\ref{eq4-11}) are the well-known Kalman-Bucy filter, and the proof of other ones are given in \citet{kurisaki2025}.

As mentioned above, the stochastic integral defining $A_t(p,q,r,\alpha;n)$ may not be interpreted as an It\^o integral, but it can be differentiated in accordance with the following rule, which is similar to It\^o's rule:\vspace{10pt}\\
{\bf Rule:} If $\alpha_{s;t}$ satisfies 
\begin{align*}
  d_t\alpha_{s;t}=p_{s;t}dt+q_{s;t}dY_t^\epsilon,
\end{align*}
then
\begin{align}
  \label{eq-derivative-rule}d_t\int_0^t \alpha_{s;t}dY_s^\epsilon=\alpha_{t;t}dY_t^\epsilon+\int_0^t p_{s;t}dY_s^\epsilon\,dt+\int_0^t q_{s;t}dY_s^\epsilon\,dY_t^\epsilon+\sigma^2q_{t;t}dt.
\end{align}
Here, the final term results from the Itô correction term $q_{t;t}dY_t\times dY_t$.\vspace{10pt}

For example, let us consider the derivative of 
\begin{align*}
  A_t(1,0,0,1;1)=\int_0^t \mu_{s;t}(dY_s^\epsilon-c\mu_{s;t}ds)=\int_0^t \mu_{s;t}dY_s^\epsilon-c\int_0^t\mu_{s;t}^2ds.
\end{align*}
By (\ref{eq4-8}) and applying (\ref{eq-derivative-rule}) yields
\begin{align*}
  d_t\int_0^t \mu_{s;t}dY_s^\epsilon=\mu_{t,t}dY_t^\epsilon+\frac{c}{\sigma^2}\int_0^t \gamma(s,t;t) dY_s^\epsilon (dY_t^\epsilon-c\mu_{t;t}dt)+c\gamma(t)dt.
\end{align*}
and
\begin{align*}
  d_t\int_0^t\mu_{s;t}^2ds=\mu_{t;t}^2dt+2\frac{c}{\sigma^2}\int_0^t\mu_{s;t}\gamma(s,t;t)ds(dY_t^\epsilon-c\mu_{t;t}dt).
\end{align*}
Thus, we obtain the equation
\begin{align*}
  dA_t(1,0,0,1;1)=&\mu_{t;t}(dY_t^\epsilon-c\mu_{t;t}dt)+\frac{c}{\sigma^2}\int_0^t \gamma(s,t;t)(dY_s-c\mu_{s;t}ds)\,(dY_t^\epsilon-c\mu_{t;t}dt)\\
  &-\frac{c^2}{\sigma^2}\int_0^t \mu_{s;t}\gamma(s,t;t)ds\,(dY_t^\epsilon-c\mu_{t;t}dt)-\frac{c^2}{\sigma^2}\int_0^t \gamma(s,t;t)^2ds\,dt+c\gamma(t)dt\\
  =&\mu_{t;t}(dY_t^\epsilon-c\mu_{t;t}dt)+\frac{c}{\sigma^2}A_t(0,1,0,1;1)(dY_t^\epsilon-c\mu_{t;t}dt)\\
  &-\frac{c^2}{\sigma^2}A_t(1,1,0,0;1)(dY_t^\epsilon-c\mu_{t;t}dt)-\frac{c^2}{\sigma^2}A_t(0,2,0,0;1)dt+c\gamma(t)dt.
\end{align*} 
In the same way, we have
\begin{align*}
  dA_t(0,1,0,1;1)=&\left\{ a-\frac{c^2\gamma(t)}{\sigma^2} \right\}\int_0^t \gamma(s,t;t)(dY_s^\epsilon-c\mu_{s;t}ds)\,dt\\
  &-\frac{c^2}{\sigma^2}\int_0^t \gamma(s,t;t)^2ds\,(dY_t^\epsilon-c\mu_{t;t}dt),\\
  =&\left\{ a-\frac{c^2\gamma(t)}{\sigma^2} \right\}A_t(0,1,0,1;1)dt-\frac{c^2}{\sigma^2}A_t(0,2,0,0;1)(dY_t^\epsilon-c\mu_{t;t}dt),\\
  dA_t(1,1,0,0;1)=&\left\{ a-\frac{c^2\gamma(t)}{\sigma^2} \right\}A_t(1,1,0,0;1)+\frac{c}{\sigma^2}\int_0^t \gamma(s,t;t)^2ds\,(dY_t^\epsilon-c\mu_{t;t}dt),\\
  =&\left\{ a-\frac{c^2\gamma(t)}{\sigma^2} \right\}dA_t(1,1,0,0;1)+\frac{c}{\sigma^2}A_t(0,2,0,0;1)(dY_t^\epsilon-c\mu_{t;t}dt),\\
  dA_t(0,2,0,0;1)=&2\left\{ a-\frac{c^2\gamma(t)}{\sigma^2} \right\}\int_0^t \gamma(s,t;t)^2ds\,dt=2\left\{ a-\frac{c^2\gamma(t)}{\sigma^2} \right\}A_t(0,2,0,0;1)dt.
\end{align*}
Therefore, $A_t(1,0,0,1;1)$ can be computed by solving the four-dimensional stochastic differential equation
\begin{align}
  dA_t(1,0,0,1;1)=&\mu_{t;t}(dY_t^\epsilon-c\mu_{t;t}dt)+\frac{c}{\sigma^2}A_t(0,1,0,1;1)(dY_t^\epsilon-c\mu_{t;t}dt)\nonumber\\
  &-\frac{c^2}{\sigma^2}A_t(1,1,0,0;1)(dY_t^\epsilon-c\mu_{t;t}dt)-\frac{c^2}{\sigma^2}A_t(0,2,0,0;1)dt+c\gamma(t)dt,\label{eq-ode-1}\\
  dA_t(0,1,0,1;1)=&\left\{ a-\frac{c^2\gamma(t)}{\sigma^2} \right\}A_t(0,1,0,1;1)dt-\frac{c^2}{\sigma^2}A_t(0,2,0,0;1)(dY_t^\epsilon-c\mu_{t;t}dt),\label{eq-ode-2}\\
  dA_t(1,1,0,0;1)=&\left\{ a-\frac{c^2\gamma(t)}{\sigma^2} \right\}dA_t(1,1,0,0;1)+\frac{c}{\sigma^2}A_t(0,2,0,0;1)(dY_t^\epsilon-c\mu_{t;t}dt),\label{eq-ode-3}\\
  dA_t(0,2,0,0;1)=&2\left\{ a-\frac{c^2\gamma(t)}{\sigma^2} \right\}A_t(0,2,0,0;1)dt,\label{eq-ode-4}
\end{align}
where $\mu_{t;t}$ and $\gamma(t)$ is given by (\ref{eq4-7}) and (\ref{eq4-11}). A numerical solution to this equation can be recursively computed using the Euler-Maruyama method.

Similarly, $A_t(p,q,r,\alpha;n)$ is eventually reduced to the case of $p=r=\alpha=0$ by repeatedly differentiating it using (\ref{eq4-8})--(\ref{eq4-10}). A closed system of equations can then be obtained due to (\ref{eq4-9}).

The algorithm can be summarized in the following steps:
\begin{enumerate}
  \item Expand the conditional expectation as in (\ref{eq4-5}).
  \item Due to the Gaussian property of $X$, express (\ref{eq4-3}) in terms of the mean $\mu_{s;t}$ and covariance $\gamma(s,u;t)$, resulting in an expression like (\ref{eq4-6}).
  \item Differentiate each term from the previous step according to (\ref{eq4-8})--(\ref{eq4-10}) and the rule (\ref{eq-derivative-rule}), and derive a finite-dimensional stochastic differential equation to calculate the term.
\end{enumerate}

\section{Numerical Studies}\label{section-numerical}
\subsection{Linear case}\label{section-numerical-linear}
In the first simulation, we set $g(x)=x$ and consider the following model 
\begin{alignat*}{2}
    & dX_t = aX_t \, dt + b \, dV_t, \qquad & X_0 = 0, \\
    & dY_t^\epsilon = (cX_t + \epsilon X_t ) \, dt + \sigma \, dW_t, \qquad & Y_0 = 0,
\end{alignat*}
In this case, the true value of the conditional expectation $E[X_t|\mathcal{Y}_t^\epsilon]$ can be calculated through the Kalman-Bucy filter, which serves as a benchmark for evaluating the accuracy of our asymptotic expansion.

\begin{figure}[t]
  \centering
  \begin{subfigure}[b]{\linewidth}
      \centering
      \includegraphics[width=0.80\linewidth]{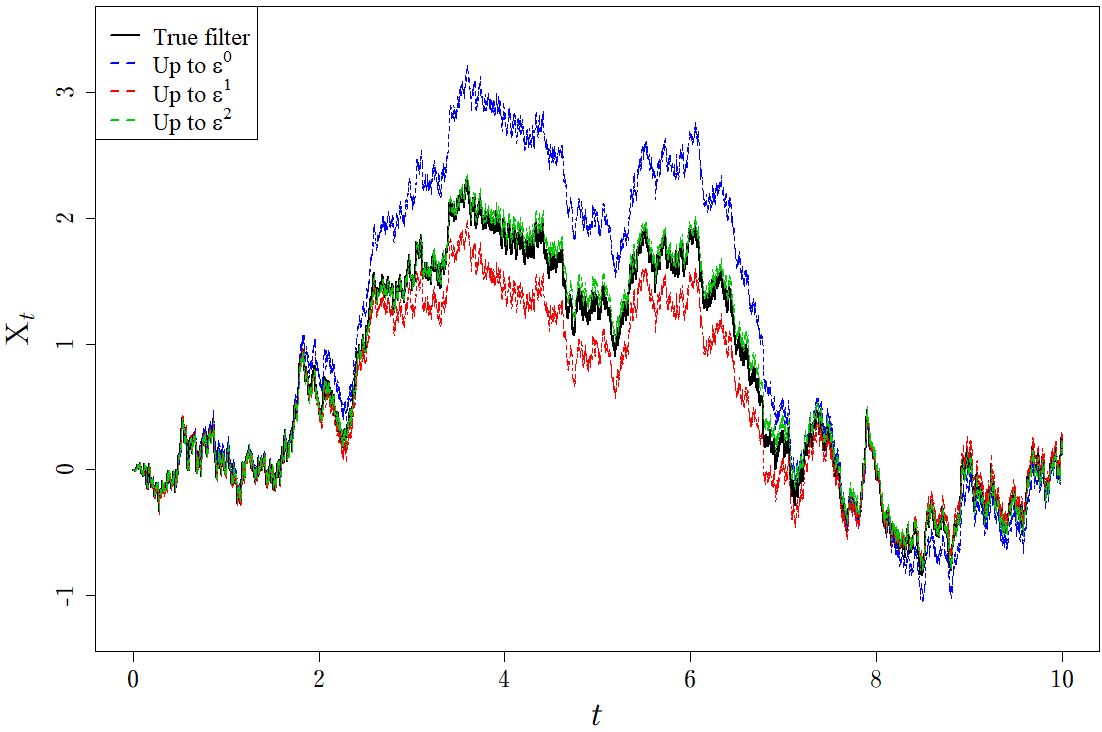}
  \end{subfigure}
  \caption{Plots of the expanded filter and the true filter.}
  \label{fig:linear-plot}
\end{figure}

We simulated a path of $X_t$ and $Y_t^\epsilon$ over the interval $0\leq t\leq 10$ with a step size of $0.001$, using the parameters
\begin{align*}
  a = -0.4, \quad b = 0.5, \quad c = 1, \quad \sigma = 0.3, \quad \epsilon = 0.2.
\end{align*}
For the simulated path, we expanded the conditional expectation $E[X_t|\mathcal{Y}_t^\epsilon]$ with respect to $\epsilon$ following the algorithm described in the previous section. Figure \ref{fig:linear-plot} shows the plots of this expansion and the true conditional expansion calculated by the Kalman-Bucy filter. presents the plots of this expansion alongside the true conditional expectation computed using the Kalman-Bucy filter. As shown in the figure, the expanded filters converge toward the true line as the order of expansion increases, demonstrating the validity of our algorithm.

\subsection{Cubic Sensor}\label{section-numerical-cubic}
In the second simulation, we consider the case of $g(x)=x^3$: 
\begin{alignat}{2}
    \label{eq-cubic-1}& dX_t = aX_t \, dt + b \, dV_t, \qquad & X_0 = 0, \\
    \label{eq-cubic-2}& dY_t^\epsilon = (cX_t + \epsilon X_t^3 ) \, dt + \sigma \, dW_t, \qquad & Y_0 = 0,
\end{alignat}
This model is referred to as the cubic sensor, which serves as a benchmark in several studies \citep{armstrong2019optimal}.

In this case, the true filter cannot be computed to directly evaluate the performance of the expansion. Instead, we generated 1,000 sample paths of $X_t$ and $Y_t$ on the interval $0 \leq t \leq 100$, aand computed the asymptotic expansion for each path. To assess the accuracy of the expansion, we evaluated the integrated squared error for each path, defined as
\begin{align*}
    \int_0^{100} (X_t - N_t^{[n],\epsilon})^2 \, dt,
\end{align*}
where 
\begin{align*}
  N_t^{[n],\epsilon}=&J_t^0(X_t) + \left\{ J_t^1(X_t) - J_t^0(X_t) J_t^1(1) \right\} \epsilon \\
  &+ \left\{ J_t^2(X_t) - J_t^0(X_t) J_t^2(1) - J_t^1(X_t) J_t^1(1) + J_t^0(X_t) J_t^1(1)^2 \right\} \epsilon^2 \\
  &+ \cdots + \sum_{k=0}^n \sum_{p=1}^{n-k} \sum_{\substack{i_1,\cdots,i_p \in \mathbb{N} \\ i_1+\cdots+i_p=n-k}} (-1)^p J_t^k(X_t) J_t^{i_1}(1) \cdots J_t^{i_p}(1) \epsilon^n
\end{align*}
is the $n$-th order approximation (see (\ref{eq4-5})). Due to the $L^2$-optimality of the conditional expectation, the true filter is expected to minimize the mean of integrated squared error.

Here, we used the following parameters:
\begin{align*}
    a = -0.4, \quad b = 0.5, \quad c = 1, \quad \sigma = 0.3, \quad \epsilon = 0.2,
\end{align*}
and simulations were performed with a step size of 0.01. $J_t^n(X_t)$ and $J_t^n(X_t) - J_t^0(X_t) J_t^n(1)$ are respectively computed by the functions \texttt{J1} and \texttt{J2} in the supplementary material.

Table \ref{table-errors} shows the minimum, median, mean, and maximum of the integrated squared error for the 1,000 simulated paths. Note that $N_t^{[0],\epsilon} = \mu_{t,t}$ corresponds to the extended Kalman filter (linear approximation).
\begin{table}[h]
  \renewcommand{\arraystretch}{1.3}
  \begin{center}
    \caption{\label{table-errors}Integrated squared errors of each filter.}
   \begin{tabular}{|l||c|c|c|}\hline
     Filter & $\mu_{t;t}$ &$N_t^{[1],\epsilon}$&$N_t^{[2],\epsilon}$ \\\hline \hline
     Min. & 8.200 &7.957&7.912\\\hline
     Median & 10.91 &10.73&11.19\\\hline
     Mean&10.98&10.76&25.01\\\hline
     Max. &16.87&35.60&9215\\\hline
   \end{tabular}
  \end{center}
 \end{table}
 We can observe the superiority of the asymptotic expansion over the extended Kalman filter in terms of the minimum and median of $N_t^{[1],\epsilon}$, but it does not perform as well in terms of the mean and maximum.

\begin{figure}[htbp]
  \centering
  \begin{subfigure}[b]{\linewidth}
      \centering
      \includegraphics[width=0.99\linewidth]{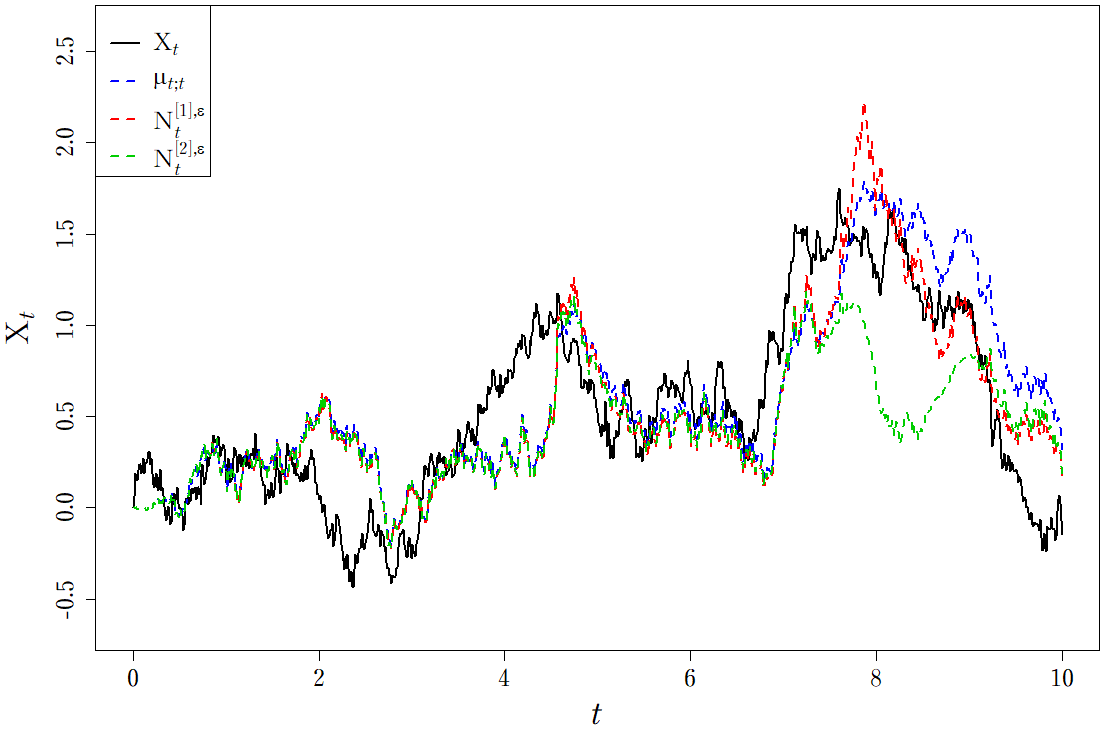}
  \end{subfigure}
  \vspace{0.2cm}

  \begin{subfigure}[b]{\linewidth}
      \centering
      \includegraphics[width=0.99\linewidth]{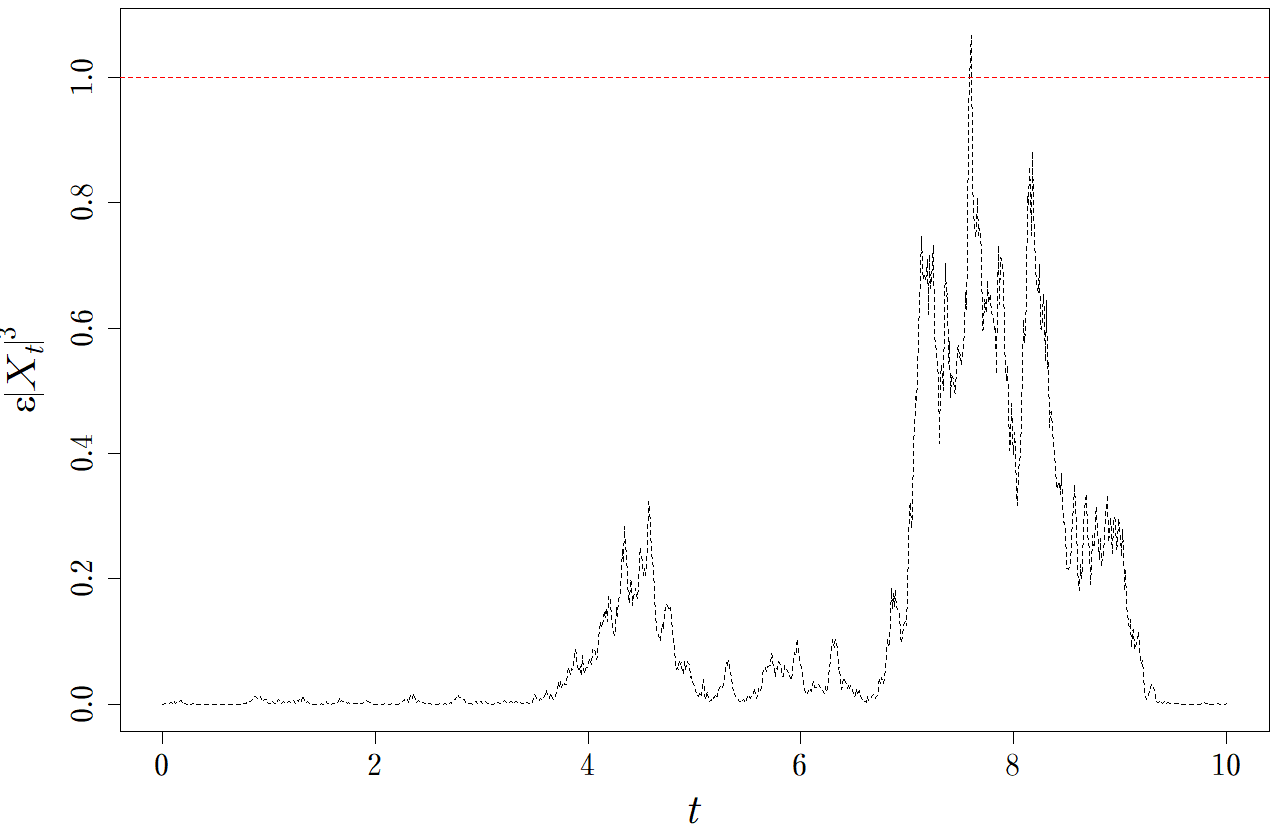}
  \end{subfigure}

  \caption{Plots of the filtering results and the perturbation term for a single simulated path.}
  \label{fig:filter-plots}
\end{figure}

This result can be explained by Figure \ref{fig:filter-plots}, which shows the plots of \(X_t\), \(\mu_{t;t}\), \(N_t^{[1],\epsilon}\), \(N_t^{[2],\epsilon}\), and \(\epsilon |X_t|^3\) for a single simulated path for the interval \(0 \leq t \leq 10\). From the figure, we observe that \(N_t^{[1],\epsilon}\) and \(N_t^{[2],\epsilon}\) improve upon the extended Kalman filter \(\mu_{t;t}\) over much of the interval, but \(N_t^{[2],\epsilon}\) begins to deviate from the true process before $t = 8$.

Looking at the lower panel, we find that this deviation aligns with the moments when the absolute value of the perturbation term \(\epsilon |X_t|^3\) exceeds 1. This behavior is expected, as our expansion is performed with respect to \(\epsilon {X_t}^3\), rather than just \(\epsilon\). Consequently, the expansion diverges when \(\epsilon |X_t|^3 > 1\). However, the upper panel suggests that this effect is not persistent, as the path of \(N_t^{[2],\epsilon}\) returns closer to the true path as soon as \(\epsilon |X_t|^3\) falls below 1. Hence, the minimum and median values in Table \ref{table-errors} capture the convergence of the asymptotic expansion when \(\epsilon |X_t|^3 < 1\), whereas the maximum and mean values reflect its divergence when \(\epsilon |X_t|^3 > 1\). 

To manage the divergence of the expansion, we propose a modification to the coefficients. Let us write \(N_t^{[n],\epsilon}\) as
\[
N_t^{[n],\epsilon} = n_t^{[0]} + n_t^{[1]}\epsilon + n_t^{[2]}\epsilon^2 + \cdots + n_t^{[n]}\epsilon^n,
\]
where the coefficients are given by:
\begin{align*}
  &n_t^{[0]} = J_t^0(X_t), \\
&n_t^{[1]} = J_t^1(X_t) - J_t^0(X_t)J_t^1(1), \quad\\
&n_t^{[2]}=J_t^2(X_t) - J_t^0(X_t) J_t^2(1) - J_t^1(X_t) J_t^1(1) + J_t^0(X_t) J_t^1(1)^2.
\end{align*}
As discussed above, the term \(n_t^{[i]}\epsilon^i\) either converges or diverges at an exponential rate, depending on the value of \(\epsilon |X_t|^3\). To suppress the divergence and ensure exponential decay, we fix a parameter \(r > 0\) and define the modified coefficients \(\tilde{n}_t^{[i]}\) as follows:
\[
\tilde{n}_t^{[0]} = n_t^{[0]}, \quad
\tilde{n}_t^{[i]} =
\begin{cases} 
n_t^{[i]} & \text{if } |n_t^{[i]}\epsilon^i| \leq r |\tilde{n}_t^{[i-1]}\epsilon^{i-1}|, \\
\displaystyle r |\tilde{n}_t^{[i-1]}\epsilon^{i-1}|\frac{n_t^{[i]}}{|n_t^{[i]}\epsilon^i|} & \text{if } |n_t^{[i]}\epsilon^i| > r |\tilde{n}_t^{[i-1]}\epsilon^{i-1}|,
\end{cases}
\]
for \(i = 1, 2, \cdots\). Using these coefficients, we define a modified filter by
\[
\tilde{N}_t^{[i],\epsilon} = \tilde{n}_t^{[0]} + \tilde{n}_t^{[1]}\epsilon + \tilde{n}_t^{[2]}\epsilon^2 + \cdots + \tilde{n}_t^{[i]}\epsilon^i.
\]
By construction, we have \(|\tilde{n}_t^{[i]}\epsilon^i| \leq r |\tilde{n}_t^{[i-1]}\epsilon^{i-1}|\), which guarantees the convergence of \(\tilde{N}_t^{[i],\epsilon}\) for all \(t \geq 0\) if \(0 < r < 1\).

\begin{figure}[h]
  \centering
      \includegraphics[width=0.70\linewidth]{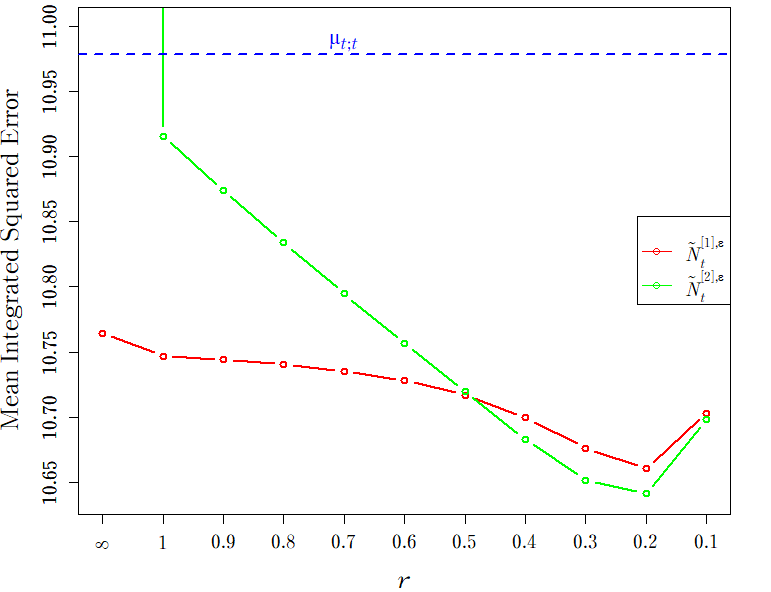}
  \caption{Mean integrated squared errors of the modified filters.}
  \label{fig:MISE-plot}
\end{figure}
\begin{figure}[h!]
  \centering
      \includegraphics[width=0.90\linewidth]{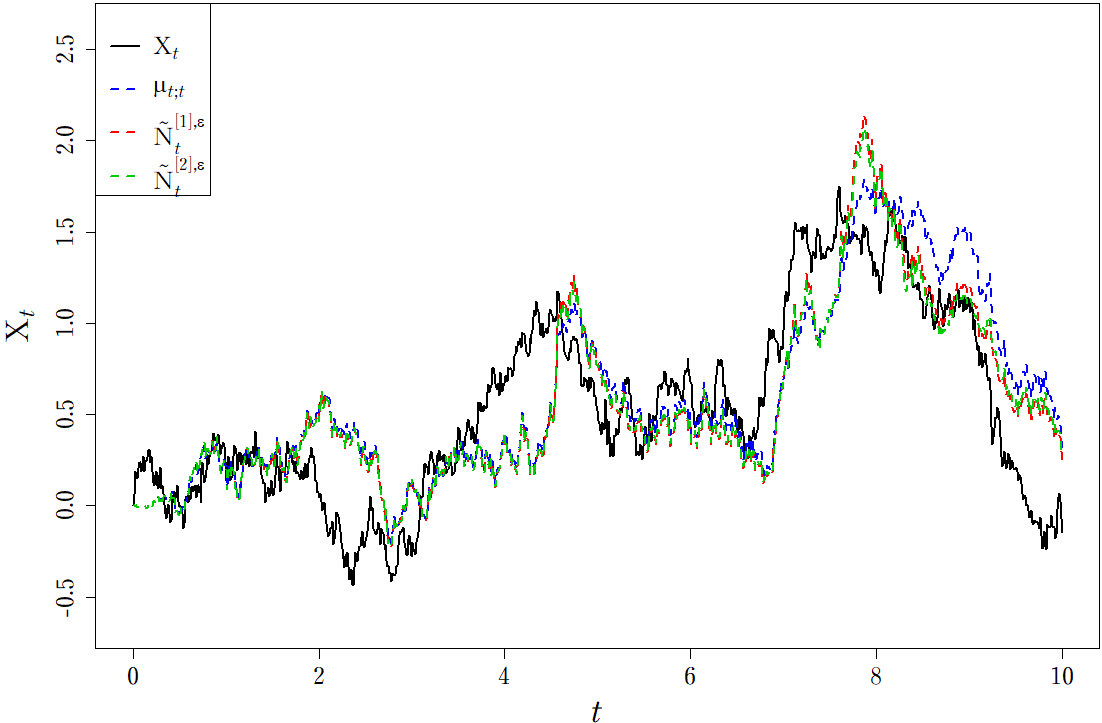}
  \caption{The modified filter of $r=0.2$ for the same simulated path as Figure \ref{fig:filter-plots}.}
  \label{fig:modified-filter-plot}
\end{figure}

Figure \ref{fig:MISE-plot} shows the mean integrated squared error of $\tilde{N}_t^{[1],\epsilon}$ and $\tilde{N}_t^{[2],\epsilon}$ for $r = 0.1, 0.2, \ldots, 1$ and $r = \infty$, based on the 1,000 simulated paths. Note that $\tilde{N}_t^{[i],\epsilon} = N_t^{[i],\epsilon}$ when $r = \infty$. From this result, we observe that the accuracy improves with the modification, and the error decreases as $r$ becomes smaller. The errors of $\tilde{N}_t^{[1],\epsilon}$ and $\tilde{N}_t^{[2],\epsilon}$ both attain their minimum at $r = 0.2$. Figure \ref{fig:modified-filter-plot} shows the modified expansions with $r = 0.2$ for the same path as in Figure \ref{fig:filter-plots}. From this figure, we observe that $\tilde{N}_t^{[1],\epsilon}$ is closer to the true path than $\mu_{t;t}$, and $\tilde{N}_t^{[2],\epsilon}$ slightly improves upon $\tilde{N}_t^{[1],\epsilon}$.

\begin{figure}[htbp]
  \centering
  \begin{subfigure}[b]{0.49\linewidth}
      \centering
      \includegraphics[width=\linewidth]{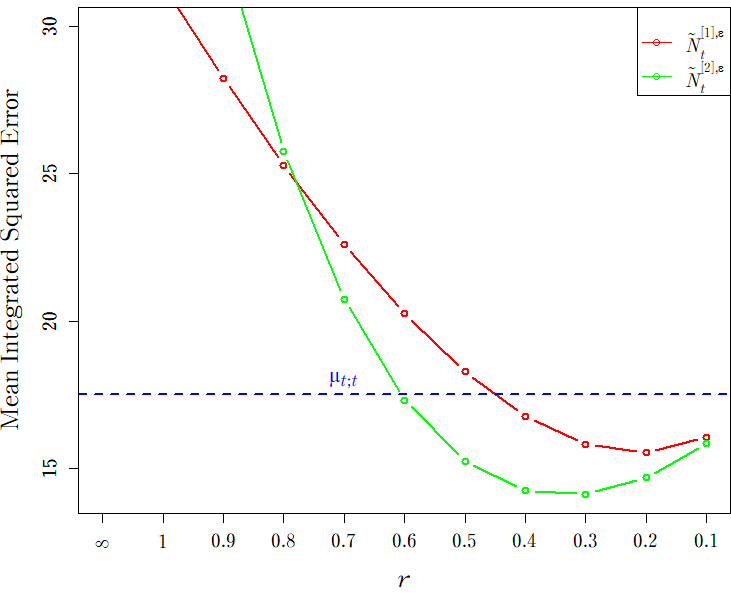}
      \caption{$\epsilon=0.8$}
      \label{fig:doctor-plot5}
  \end{subfigure}
  \hfill
  \begin{subfigure}[b]{0.49\linewidth}
      \centering
      \includegraphics[width=\linewidth]{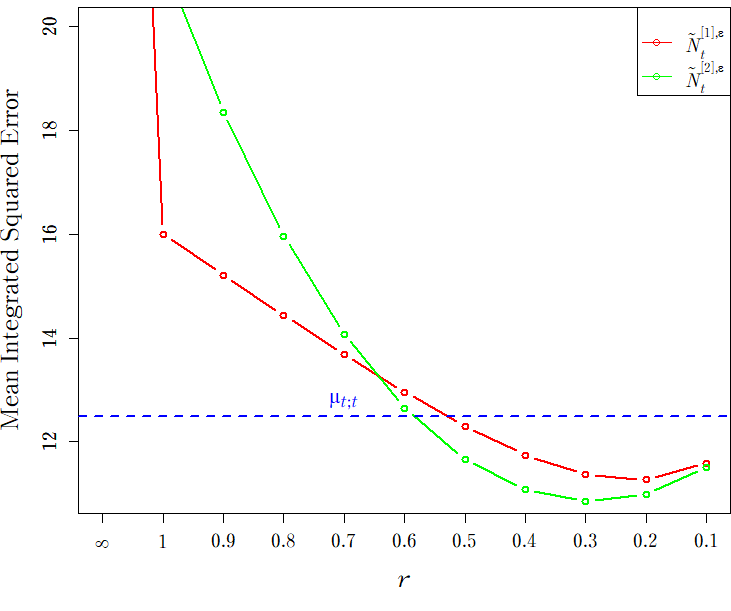}
      \caption{$\epsilon=0.5$}
      \label{fig:doctor-plot6}
  \end{subfigure}

  \vspace{0.5cm} 

  \begin{subfigure}[b]{0.49\linewidth}
      \centering
      \includegraphics[width=\linewidth]{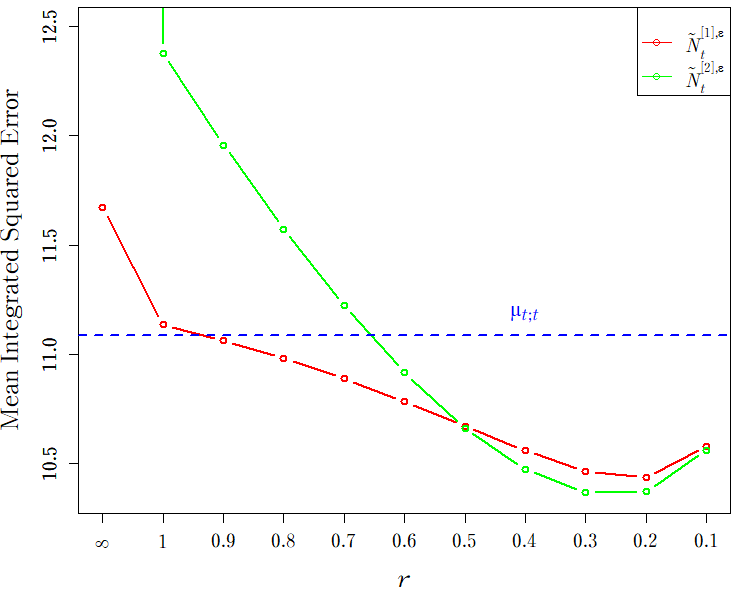}
      \caption{$\epsilon=0.3$}
      \label{fig:doctor-plot7}
  \end{subfigure}
  \hfill
  \begin{subfigure}[b]{0.49\linewidth}
      \centering
      \includegraphics[width=\linewidth]{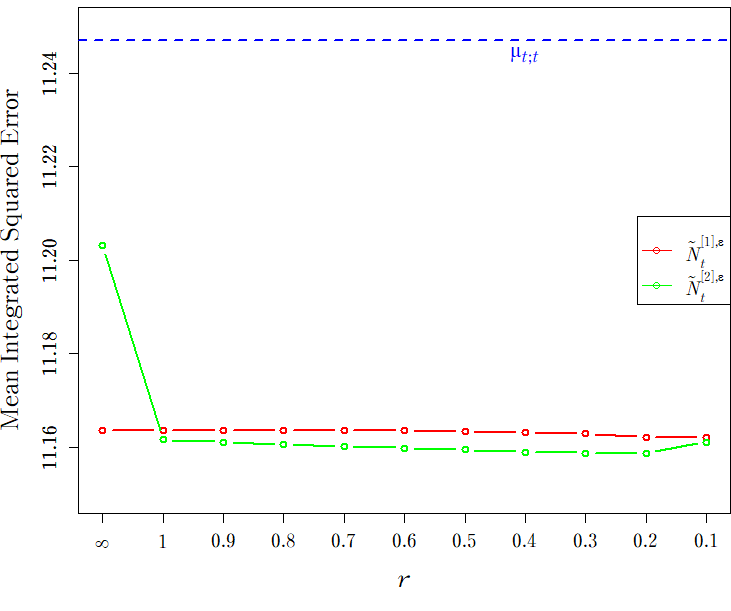}
      \caption{$\epsilon=0.1$}
      \label{fig:doctor-plot8}
  \end{subfigure}

  \caption{Mean integrated squared errors for $\epsilon=0.8,0.5,0.3,0.1$.}
  \label{fig:plots-for-other-epsilons}
\end{figure}

The optimal value of $r$ coincides with $\epsilon = 0.2$ in this case, but they are not necessarily related. In fact, as shown in Figure \ref{fig:plots-for-other-epsilons}, the error is minimized around $r = 0.3$ regardless of the value of $\epsilon$. For each $\epsilon$, we observe that $\tilde{N}_t^{[1],\epsilon}$ and $\tilde{N}_t^{[2],\epsilon}$ can improve upon the linear approximation for sufficiently small $r$, and $\tilde{N}_t^{[2],\epsilon}$ attains a smaller error than $\tilde{N}_t^{[1],\epsilon}$.

In summary, although the raw asymptotic expansion does not necessarily improve the accuracy of the filter over the entire time horizon due to the divergence of the perturbation term, we can control this divergence by choosing an appropriate parameter $r$ and modifying the coefficients of the expansion. With this modification, we observe an overall improvement in the filter's accuracy. While it is difficult to theoretically determine the optimal value of the parameter $r$, a relatively good value of $r$ can be obtained by conducting simulations in advance. 

\subsection{Expansion of the density}\label{section-numerical-density}
So far, we have focused on the expansion of the conditional expectation. However, the same approach can also be applied to the conditional density by considering the expansion of the conditional characteristic function $E[e^{iu X_t}|\mathcal{Y}_t]$.  In this subsection, we consider the first-order expansion of this characteristic function.

As in equation~\eqref{eq4-5}, we obtain the following expansion:
\begin{align*}
  E[e^{iu X_t}|\mathcal{Y}_t]=J_t^0(e^{iuX_t})+\left\{ J_t^1(e^{iuX_t})-J_t^0(e^{iuX_t})J_t^1(1) \right\}\epsilon+O(\epsilon^2).
\end{align*}
Since $X_t$ is Gaussian with mean $\mu_{t;t}$ and $\gamma(t)$ under $\tilde{E}_t$, we have 
\begin{align*}
  J_t^0(e^{iuX_t})=\tilde{E}_t[e^{iuX_t}]=e^{iu\mu_{t;t}-\frac{1}{2}\gamma(t)u^2}.
\end{align*}

For $J_t^1(e^{iuX_t})$, we proceed as follows:
\begin{align}
  J_t^1(X_t)&=\tilde{E}_t\left[ e^{iuX_t}\int_0^t X_s^3 (dY_s^\epsilon - cX_sds) \right]\nonumber\\
  &=e^{iu\mu_{t,t}-\frac{1}{2}\gamma(t)u^2}\tilde{E}_t\left[e^{iu(X_t+\mu_{t;t})+\frac{1}{2}\gamma(t)u^2} \int_0^t X_s^3 (dY_s^\epsilon - cX_sds) \right]\nonumber\\
  &=e^{iu\mu_{t,t}-\frac{1}{2}\gamma(t)u^2}\sum_{k=0}^\infty\frac{1}{k!}\tilde{E}_t\left[H_k(X_t-\mu_{t;t};\gamma(t)) \int_0^t X_s^3 (dY_s^\epsilon - cX_sds) \right](iu)^k,\label{eq-characteristic}
\end{align}
where
\begin{align*}
  H_k(x;\gamma(t))=\left(\frac{\partial^k}{\partial y^k}e^{yx-\frac{1}{2}\gamma(t)y^2}\right)_{y=0}
\end{align*}
are Hermite polynomials. 

Due to the Gaussianity of $X_t$ (under $\tilde{E}_t$) and the orthogonality of Hermite polynomials (or equivalently, the orthogonality of Wiener chaos spaces), we have
\begin{align*}
  \tilde{E}_t\left[H_k(X_t-\mu_{t;t};\gamma(t)) \int_0^t X_s^3 (dY_s^\epsilon - cX_sds) \right]=0~~(k\geq 5).
\end{align*}
Therefore, in the expansion~\eqref{eq-characteristic}, it suffices to consider terms up to $k\leq 4$. Alternatively, the first-order approximation of the conditional density can also be obtained via an Edgeworth expansion using cumulants up to order four. 

\begin{figure}[H]
  \centering
  \begin{subfigure}[b]{\linewidth}
      \centering
      \includegraphics[width=0.70\linewidth]{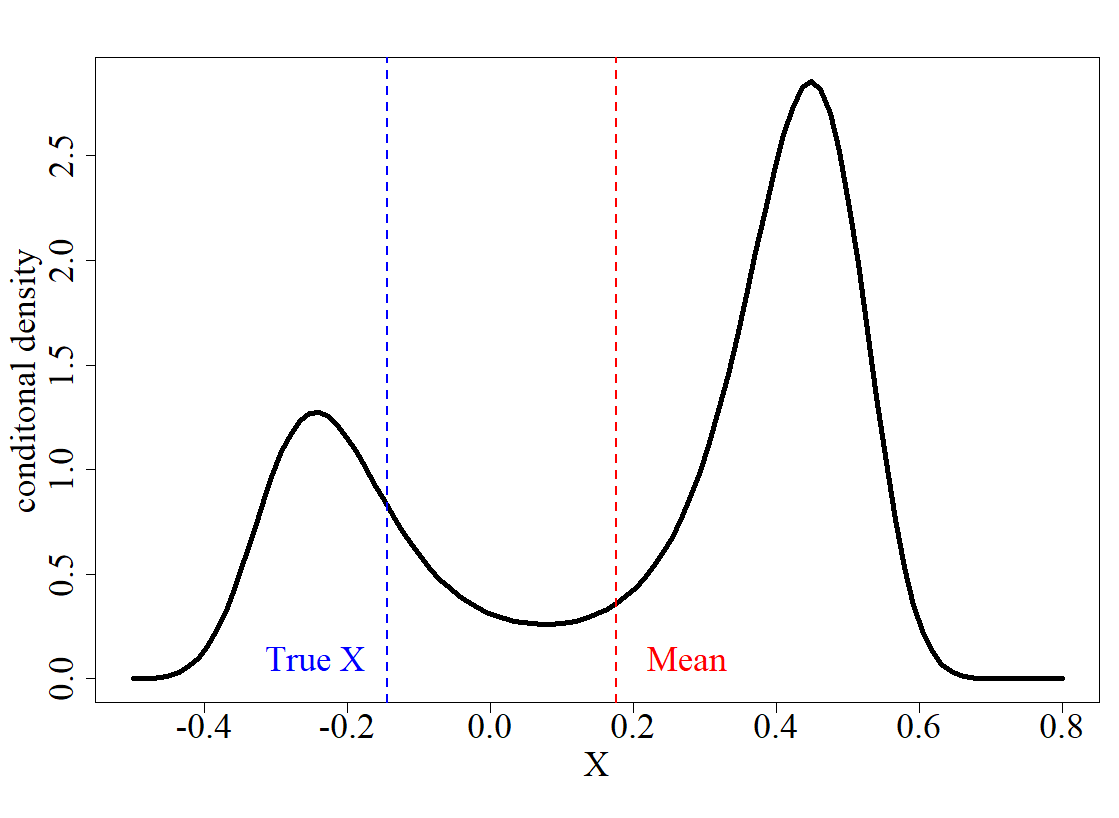}
  \end{subfigure}
  \caption{First order approximation of the conditional density of $X_{10}$.}
  \label{fig:density-plot}
\end{figure}

Figure~\ref{fig:density-plot} shows the first-order approximation of the conditional density of $X_t$ at $t=10$ corresponding to the simulated path shown in Figure~\ref{fig:filter-plots}. We observe that the approximated density exhibits two distinct modes. The conditional mean lies between these peaks, while the true value is located closer to the left mode.

\section{Discussion}\label{section-discussion}
In our approach, the computation of the asymptotic expansion is reduced to solving ordinary differential equations, such as (\ref{eq-ode-1})--(\ref{eq-ode-4}). This yields a significant computational advantage over particle filters and PDE-based methods. In particular, it suggests that our method has the potential to be applied in high-dimensional settings, such as data assimilation, where these existing approaches become computationally prohibitive and linearization methods are typically employed. Furthermore, the results in Figure~\ref{fig:density-plot} highlight the advantage of our method in capturing complex features of the hidden process, such as multi-modality, that cannot be handled by standard Gaussian or linear approximations.

In this paper, we have provided numerical results only for the perturbed linear model. The simulation for the original nonlinear model has not yet been conducted, primarily due to the complexity of its implementation. Nevertheless, the reformulation part essentially amounts to a Taylor expansion, and the subsequent procedures closely follow those of the perturbed linear case. Moreover, we expect that the modification of the coefficients will not be required in the nonlinear case. This is because the polynomial terms in equation~(\ref{eq-expansion-h}) arise as higher-order terms in the expansion of $h(X_t^\epsilon)$, and their smallness is therefore expected.

\section{Disclosure statement}\label{disclosure-statement}
The authors declare that there are no conflicts of interest regarding the publication of this paper.
\section{Data Availability Statement}
All data and code used in this study were generated by the author. No external datasets were used.

\phantomsection\label{supplementary-material}
\bigskip

\begin{center}

{\large\bf SUPPLEMENTARY MATERIAL}

\end{center}

\begin{description}
\item[R archive:]
Compressed archive containing all R scripts used for implementation and simulation. The archive includes:
\begin{itemize}
  \item \texttt{funcs.R} – implementation of functions necessarily for asymptotic expansion. In particular, \texttt{J1(i, j, n, a, b, sigma, obs, delta)} and \texttt{J2(i, j, n, a, b, sigma, obs, delta)} compute $J_t^n(X^i)$ and $J_t^n(X^i)-J_t^0(X^i)J_t^0(1)$ defined by (\ref{eq4-3}), respectively, under $g(x)=x^j$ and $c=1$.
  \item \texttt{Avec\_solve.cpp} – implementation of a C++ function to compute $A_t(p,q,r,\alpha;n)$ in (\ref{eq-Avec}).
  \item \texttt{sim\_linear.R} – code for reproducing results in Section \ref{section-numerical-linear}.
  \item \texttt{sim\_cubic.R} – code for reproducing results in Section \ref{section-numerical-cubic}. The script implements a single-path simulation, but the 1000-path average results can be reproduced by running the simulation repeatedly.  
\end{itemize}
\end{description}

\bibliographystyle{abbrvnat}

  \bibliography{paper3-arxiv.bbl}

\end{document}